\documentclass[preprint]{elsarticle}

\usepackage{graphics}
\usepackage{eufrak}
\usepackage{amsmath,amssymb}
\usepackage{amsmath,color}
\usepackage{graphicx, hyperref}
\usepackage{psfrag}
\usepackage{textcomp}
\usepackage[usenames,dvipsnames]{xcolor}
\usepackage{bm}

\biboptions{sort&compress}

\begin{document}
\title{Gravitational Entropy of Hayward Black Hole}

\author{Hideo Iguchi}
\ead{iguchi.hideo@nihon-u.ac.jp}

\address{Laboratory of Physics, College of Science and Technology, Nihon University, 
274-8501 Narashinodai, Funabashi, Chiba, Japan}

\begin{abstract}
We analyze the gravitational entropy defined by the Weyl curvature for the Hayward black hole, which is one of the regular black holes without singularity in the event horizon. 
Using the definition by the ratio of the Weyl curvature scalar and the Kretschmann scalar as the entropy measure, we evaluate the gravitational entropy on the outer and inner horizons. 
We derive an expression for the curvature tensor of the Hayward black hole that is as independent as possible of the parameter added to the black hole and give an equation for the entropy measure explicitly independent of that parameter. 
The gravitational entropy density used in previous studies presents questions from the standpoint of mathematical rigor while we discuss possible improvements in its definition. 
We compare the results for the Hayward black hole with the analysis for the Reisner-Nordst\"{o}m black hole and discuss the effect of singularity resolution on gravitational entropy. 
\end{abstract}
\begin{keyword}
Black hole \sep Regular black hole \sep Gravitational entropy \sep Black hole thermodynamics \sep Weyl curvature hypothesis
\end{keyword}

\date{\today}

\maketitle

\section{Introduction}
\label{sec:Introduction}
Remarkable progress has been made in the observational study of black holes recently.
For example, gravitational waves observed by laser interferometry are considered sufficient evidence to believe in the existence of binary black holes \cite{LIGOScientific:2016aoc}.
Imaging of a supermassive black hole by EHT has identified a photon sphere and black hole shadow that are believed to exist around the event horizon \cite{EventHorizonTelescope:2019dse}.
If observation techniques continue to develop, the quality of information obtained from observations of black holes is improved while the amount of information will increase enormously.
With these developments, the theoretical study of black holes will become increasingly important.

One of the milestones in the theoretical study of black holes is the singularity theorem by Penrose \cite{Penrose:1964wq,Hawking:1970zqf}.
According to the singularity theorem, there is always a spacetime singularity in the event horizon of a black hole.
Since classical physical laws break down at a spacetime singularity, we assume that some quantum effect eliminates the singularity.
Although the theory of quantum gravity, which combines general relativity and quantum mechanics without contradiction, has not yet been completed, several researchers have proposed models of regular black holes that resolve singularities in the event horizon in the scope of classical general relativity since Bardeen's first proposal \cite{bardeen1968non}.
{
Ay\'on-Beato and Garc\'ia  \cite{Ayon-Beato:1998hmi,Ayon-Beato:1999qin,Ayon-Beato:1999kuh,Ayon-Beato:2000mjt,Ayon-Beato:2004ywd} showed that the regular black hole models can be obtained as a solution of the Einstein equation coupled with a physical source of an electric or magnetic monopole in nonlinear electrodynamics.
Dymnikova \cite{Dymnikova:1992ux} found an electric  spherically symmetric regular black hole solution with a de Sitter core.
The existence of regular black hole solutions with an electronic charge seems to contradict
the no-go theorem by Bronikov \cite{Bronnikov:1979ex}. 
One can resolve this apparent contradiction by considering that in a regular solution there are different Lagrangian functions for separete regions \cite{Bronnikov:2000vy}.
In the review article \cite{Bronnikov:2022ofk}, Bronnikov introduces this issue. 
Fan and Wang \cite{Fan2016} presented a procedure for constructing regular black hole solutions in nonlinear electrodynamics and obtained a broad class of spherically symmetric black hole solutions that include the Bardeen \cite{bardeen1968non} and the Hayward black hole \cite{Hayward2006}. 
}
For the history of research on regular black holes, the references \cite{Ansoldi:2008jw}, \cite{Lemos:2011dq} and \cite{Maeda2022} would be instructive.
We expect that the features of regular black holes appear not only inside the event horizon where there is no singularity but also at the event horizon and outside of it. 
Observations of black holes by gravitational waves and radio waves will provide knowledge about the resolution of spacetime singularities due to quantum effects and the theory of quantum gravity.
Based on these expectations, many studies have been conducted in the last decade on various physical phenomena, for example, particle motion around regular black holes, light spheres, black hole shadows, etc.\footnote{The number of references is too large to introduce each one. 
{
See, for example, \cite{Isomura:2023oqf} and reference therein.}
}

There is supposed to be a significant disruption of the homogeneity and isotropy of spacetime at the singularity of a black hole. 
This information appears in the Weyl curvature of spacetime. The Weyl curvature hypothesis~\cite{penrose1979singularities} proposed by Penrose argues that gravitational entropy is defined by the Weyl curvature of spacetime since the Weyl curvature carries the degrees of freedom of the gravitational field. 
In this hypothesis, since the Weyl curvature is zero at the initial singularity of the homogeneous isotropic universe model, the gravitational entropy is zero or very small at the initial singularity of the universe, and it has increased due to the structure formation associated with the evolution of the universe. 
Thus, although at first glance the universe appears to have evolved from disorder to order with the structure formation, we can consider that the second law of thermodynamics does not break down due to the increase in gravitational entropy.

To investigate whether the Weyl curvature tensor is a direct measure of entropy, Rudjord {\it et al.}~\cite{Rudjord2008} have studied an evaluation of gravitational entropy based on Weyl curvature in black hole spacetime. 
They defined an entropy measure using the ratio of the Weyl curvature scalar to the Kretschmann scalar. 
They also proposed how to calculate the gravitational entropy by a surface integral over the event horizon of the black hole of the entropy measure. Gr{\o}n~\cite{Gron:2012} has provided a further discussion on this issue. Several researchers have calculated gravitational entropy in various black hole and wormhole spacetimes using this entropy measure~\cite{Romero2012,Perez:2014oea,Guha:2019fun,deCLima:2020rvr,Chakraborty2022}. 
Clifton et al.~\cite{Clifton2013} derived the effective energy density of a free gravity field using the Bel-Robinson tensor and proposed a method to calculate entropy using relativistic thermodynamics. 
Gregoris and Ong~\cite{Gregoris2022} presented a way to evaluate the gravitational entropy density from the Newman-Penrose scalar and the first-order frame derivative of the Weyl curvature tensor.

In this paper, we evaluate the gravitational entropy of a regular black hole and discuss how the resolution of the spacetime singularity by quantum gravitational effects affects the gravitational entropy of the black hole, i.e., the degree of freedom of the gravitational field inside the event horizon. 
One can regularize the central singularity of a black hole by replacing it with a de Sitter-like core. Since the Weyl curvature is zero in de Sitter spacetime, the gravitational entropy would be zero or very small inside a regular black hole. 
As for a similar analysis,  P\'{e}res {\it et al.}~\cite{Perez:2014oea} have applied the entropy measure of Rudjord {\it et al.} to the regular black hole model given by  Mbonye-Kazanas~\cite{Mbonye:2005im}. 
Here we adopt the one derived by Hayward~\cite{Hayward2006} as the regular black hole spacetime and analyze it using the entropy measure defined by Rudjord {\it et al.}~\cite{Rudjord2008} as in a similar way as P\'{e}res {\it et al.}~\cite{Perez:2014oea}.

In Section \ref{sec:GE} we briefly describe the method proposed by Rudjord et al. for evaluating gravitational entropy in black hole spacetime.
In the next section \ref{section:Hayward} we describe the Hayward black hole and then analyze the gravitational entropy in the spacetime.
In Section \ref{sec:Discussion} we discuss the results of our analysis and provide a summary of this work.
We adopt the Planck unit $ c = G = \hbar = k_\mathrm{B} = 1$.

\section{Gravitational Entropy of Black Holes} 
\label{sec:GE}

This section describes the method of evaluating gravitational entropy used by Rudjord {\it et al.}~\cite{Rudjord2008}. 
The Bekenstein-Hawing entropy of a black hole is a quantity of geometric origin and is proportional to the event horizon area~\cite{Bekenstein:1973ur,Bardeen:1973gs,Hawking:1975vcx,Hawking:1976de}. 
Since the gravitational entropy is a geometric quantity evaluated from the Weyl curvature, we can assume that the gravitational entropy reproduces the Bekenstein-Hawing entropy. 
Of course, this is an assumption, and it may not be necessary to explain all black hole entropy by gravitational entropy. 
However, it is considered one of the most significant assumptions to consider in constructing a method for evaluating the gravitational entropy of a black hole~\cite{Rudjord2008,Clifton2013,Gregoris2022}.

If the gravitational entropy of a black hole is proportional to the event horizon area, it can be assumed to be the surface integral of some 3-dimensional vector field $\bm{\Psi} $ on the event horizon $\sigma$,
\begin{equation}
 S = k_\mathrm{S} \int_\sigma \bm{\Psi} \cdot \bm{d\sigma} ,
 \label{eq:def_S}
\end{equation}
where $k_\mathrm{S} $ is a constant.
In the case of spherically symmetric static spacetime, $\bm{\Psi}$ can be written as
\begin{equation}
 \bm{\Psi} = P \bm{e_r},
 \label{eq:Psi}
\end{equation}
where $\bm{e_r}$ is a unit radial vector.
Following the Weyl curvature hypothesis, the scalar quantity $P$ is derived from the Weyl curvature tensor.
Using the Weyl curvature scalar $W$ and the Kretschmann scalar $K$, Rudjord {\it et al.}~\cite{Rudjord2008} assumed that P is given by
\begin{equation}
 P^2 = \frac{W}{K} = \frac{C^{abcd}C_{abcd}}{R^{abcd}R_{abcd}},
 \label{eq:ent_measure}
\end{equation}
where $C_{abcd}$ is Weyl curvature tensor and $R_{abcd}$ is Riemann curvature tensor.
In the case of the Schwarzschild black hole, the Weyl and Kretschmann scalars are equal,
\begin{equation}
 W = K = \frac{12 r_\mathrm{s}^2}{r^6},
\end{equation}
where $r_\mathrm{s}$ is the Schwarzschild horizon radius, so  the entropy measure is obtained as $P^2=1$.  
Therefore, by choosing the constant $k_\mathrm{S}$ appropriately, $S$ can be made consistent with the Bekenstein-Hawking entropy $S_\mathrm{BH} = \pi r_\mathrm{s}^2$.
In \cite{Rudjord2008}, this gravitational entropy evaluation method was applied to Reisner-Nordstr\"{o}m-de Sitter black hole spacetime, and it was confirmed that the value of S is smaller than the Bekenstein-Hawking entropy for Reisner-Nordst\"{o}m and Schwarzschild-de Sitter black holes.
Romero {\it et al.}~\cite{Romero2012} reanalyzed the gravitational entropy of the Reisner-Nordstr\"{o}m black hole and also pointed out that $P^2=1$ for the Kerr black hole.

Using Gauss's divergence theorem, Eq. (\ref{eq:def_S}) can be rewritten as a volume integral as,
\begin{equation}
 S  
    = k_\mathrm{S} \int_V \nabla \cdot \bm{\Psi} dV .
\end{equation}
From this equation, it appears that the entropy density can be defined as
\begin{equation}
 s = k_\mathrm{S} \nabla \cdot \bm{\Psi}.
 \label{eq:ent_density}
\end{equation}
Since the analysis of Rudjord {\it et al.}~\cite{Rudjord2008} confirmed that $s$ can be imaginary or negative, they demanded the following equation as the entropy density from physical requirements,
\begin{equation}
 s = k_\mathrm{S} |\nabla \cdot \bm{\Psi}|.
 \label{eq:abs_s}
\end{equation}
One seems to be careful in interpreting $s$ defined in this way as an entropy density.
The result of volume integration using Eq. (\ref{eq:abs_s}) does not necessarily agree with the surface integral in Eq. (\ref{eq:def_S}).
The reason why $s$ in Eq. (\ref{eq:abs_s}) becomes imaginary is that the constant time coordinate surface in black hole spacetime is timelike inside the event horizon.
Rudjord {\it et al.} have shown that the gravitational entropy calculated in (\ref{eq:def_S}) is zero for an extremal Reisner-Nordst\"{o}m black hole.
In this case, the entropy density must always be zero or become negative in a region.
We will discuss this issue again in the next section.

\section{Hayward Black Hole and Gravitational Entropy}
\label{section:Hayward}
Since the proposal by Bardeen \cite{bardeen1968non}, many researchers have proposed and studied models of regular black holes with no spacetime singularity.
The Hayward black hole~\cite{Hayward2006} is one of the static spherically symmetric solutions of these regular black holes.
{
Fan and Wang  \cite{Fan2016} specifically wrote down the expression for the Lagrangian density of nonlinear electromagnetic fields and obtained a spherically symmetric solution with a magnetic charge, which includes the Hayward black hole.
It has been pointed out that one should set appropriately the integration constants of spherically symmetric solution coupled with nonlinear electromagnetic fields to avoid central curvature singularity \cite{Chinaglia:2017uqd,Chinaglia:2018gvf}.
As for the general solution obtained by Fan and Wang, Toshmatov, Stuchl\'ik and Ahmedov \cite{Toshmatov:2018cks} gave a comment that the gravitational mass should be equal to the electromagnetically induced mass to regularize the black hole spacetime.
}

The properties of Hayward black hole have been analyzed extensively.
For example, several researchers analyzed the thermodynamic properties of pristine (unmodified and unextended) Hayward black holes in references~\cite{Tharanath:2014naa,{Fan2016},Maluf2018,Molina:2021hgx,Fathi2021,IlichKruglov:2021pdw}.\footnote{Although there are many other studies that have analyzed the thermodynamic properties of modified and extended Hayward black hole models, we do not list them here one by one.}
{
In these analyses, the entropy of the black hole is given by the area of event horizon. 
The temperature of the black hole is either the Hawking temperature determined by the surface gravity at the horizon or the temperature calculated based on the first law of thermodynamics.
It has been pointed out that these temperatures are consistent for vacuum black hole solutions such as the Schwarzschild black hole, while they are not consistent for regular black holes.
Ma and Zhou \cite{Ma2014,Ma:2015gpa} argued that an extra factor is necessary for the first law of thermodynamics for regular black holes to solve this problem. 
In \cite{Maluf2018}, an analysis of the temperature and the first law of thermodynamics of the Hayward black hole was performed based on the assertions by Ma and Zhou.
Zhang and Gao \cite{Zhang:2016ilt} improved on Rasheed's formulation \cite{Rasheed:1997ns} and derived a generalized first law of thermodynamics that includes additional terms from the Lagrangian of a nonlinear gauge field coupled to gravity. 
 }

\begin{figure}[ht]
 \begin{center}
   \includegraphics[width=8cm]{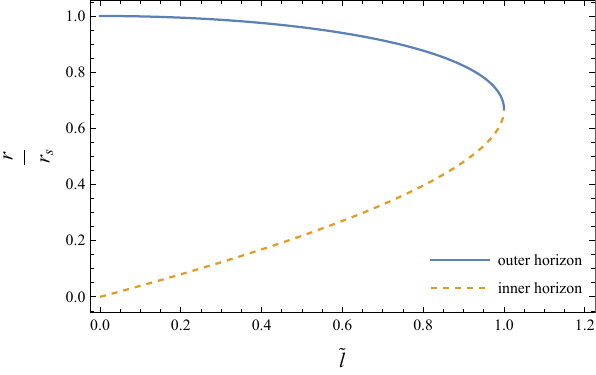}
  \end{center}
  \caption{Plot of outer (blue solid) and inner (orange dashed) horizon radius as a function of Hayward parameter $l$ normalized by the extremal value $l_0 = \frac{2}{3 \sqrt{3}} r_\mathrm{s}$.}
 \label{fig:rhtol0}
\end{figure}

The Hayward black hole spacetime metric is given by
\begin{equation}
 ds^2 = - f(r) dt^2 + \frac{dr^2}{f(r)} + r^2 d \theta^2 + r^2 \sin^2 \theta d \phi^2,
 \label{eq:metric}
\end{equation}
in which
\begin{equation}
 f(r) = 1 - \frac{r_\mathrm{s} r^2}{r^3 + r_\mathrm{s} l^2},
\end{equation}
where $r_s = 2 M$ is the horizon radius of the Schwarzschild black hole of mass $M$ and $l$ is Hayward's length parameter whose range is $0 \le l < \infty$.
{
In the original work by Hayward, this parameter is assumed to be of the order of the Planck length.
Also, based on the references \cite{Fan2016,Fan:2016rih,Bronnikov:2000vy,Bronnikov:2017tnz,Toshmatov:2018cks}, it can be considered to be related to magnetic charges.
}
The Hayward metric coincides with the Schwarzschild metric at $l=0$.
The causal structure of the Hayward black hole has an outer event horizon and an inner Cauchy horizon and is similar to the Reisner-Nordst\"{o}m black hole except that the central singularity is replaced by a regular center.
These two horizons are situated at the roots of the equation $f(r)=0$, and the explicit expressions for these roots,
\begin{align}
 \frac{r_{+}}{r_s} & =   \frac{1}{3} + \frac{2}{3} \cos \left( \frac{1}{3} \cos^{-1} \left( 1 - \frac{27 l^2}{2 r_s^2} \right) \right), \\
 \frac{r_{-}}{r_s} & =   \frac{1}{3} - \frac{2}{3} \cos \left( \frac{1}{3} \cos^{-1} \left( 1 - \frac{27 l^2}{2 r_s^2}  \right)  + \frac{\pi}{3} \right)
\end{align}
are presented in a research \cite{Chiba:2017nml} that analyzed geodesics in the Hayward black hole, where $r_{+}$ and $r_{-}$ are the outer and inner horizon radii, respectively.
The outer event horizon radius and inner Cauchy horizon radius are plotted in Fig. \ref{fig:rhtol0} as a function of the normalized Hayward parameter $\tilde{l} = \frac{l}{l_0}$, where $l_0 = \frac{2}{3 \sqrt{3}} r_\mathrm{s}$ is the limit at which the outer and inner horizon radii coincide.

\begin{figure}[ht]
 \begin{center}
   \includegraphics[width=8cm]{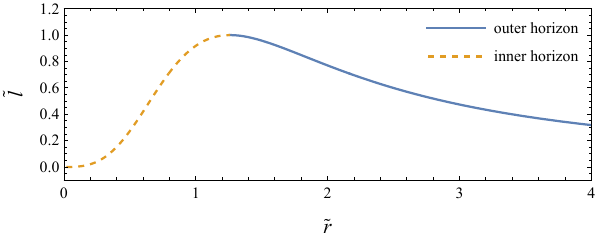}
  \end{center}
  \caption{The horizontal axis is $\tilde{r}$ of inner and outer horizons and the vertical axis is $\tilde{l}$. The blue solid line corresponds to the outer horizon and the orange dashed line to the inner horizon.}
 \label{fig:l0tort}
\end{figure}

Here we calculate the gravitational entropy of the Hayward black hole based on the entropy measure of Rudjord {\it et al}.
In this calculation, the dimensionless radial coordinate $\tilde{r} = \frac{r}{\left( l^2  r_\mathrm{s}\right)^{\frac{1}{3}}} $ is used. 
Taking the limit $l \rightarrow 0$, we get $\tilde{r} \rightarrow \infty$, so the Schwarzschild black hole is degenerate to infinity in this coordinate.
The horizon radius of the extreme Hayward black hole is $\tilde{r}_{\pm}=\sqrt[3]{2}$.
Figure \ref{fig:l0tort} shows the relation between the horizon radius and $\tilde{l}$. It is essentially the same as Fig. \ref{fig:rhtol0},  but will be useful when considering curvature scalars and entropy measures below.

\begin{figure}[ht]
 \begin{center}
   \includegraphics[width=8cm]{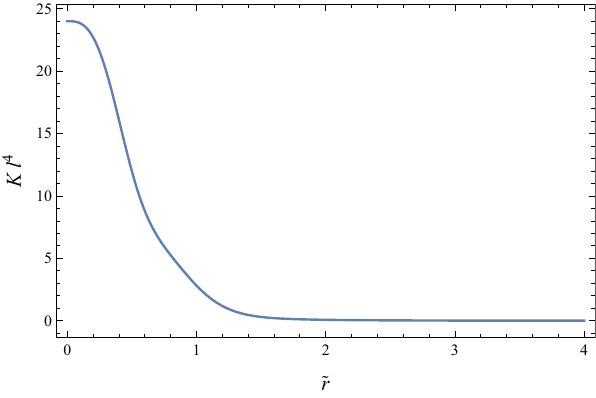}
  \end{center}
  \caption{Plot of Kretschmann scalar $K$ as a function of $\tilde{r}$. $K$ becomes large but finite aroud the center $\tilde{r}=0$.}
 \label{fig:Ktort}
\end{figure}

The Kretschmann scalar of the Hayward black hole can be calculated with some efforts and is obtained as
\begin{equation}
 K  =  \frac{12 \left( \tilde{r}^{12}  - 4 \tilde{r}^9 + 18 \tilde{r}^6 - 2 \tilde{r}^3 + 2 \right)}{l^4  \left(\tilde{r}^3 + 1 \right)^6 }.
 \label{eq:KS}
\end{equation}
In this expression, the dependence of the Kretschmann scalar on the Hayward parameter $l$ is renormalized to the radial coordinate $\tilde{r}$, except for the factor $l^{-4}$, which scales the whole.
The plot of the Kretschmann scalar is shown in Fig. \ref{fig:Ktort}.
As expected, its value remains finite and is maximal at the center.
As can be seen from Fig. \ref{fig:l0tort}, the outer horizon is located at $\tilde{r } > \sqrt[3]{2}$, and thus the location where the Kretschmann scalar becomes large is inside the outer horizon, especially localized near the center.

\begin{figure}[ht]
 \begin{center}
   \includegraphics[width=8cm]{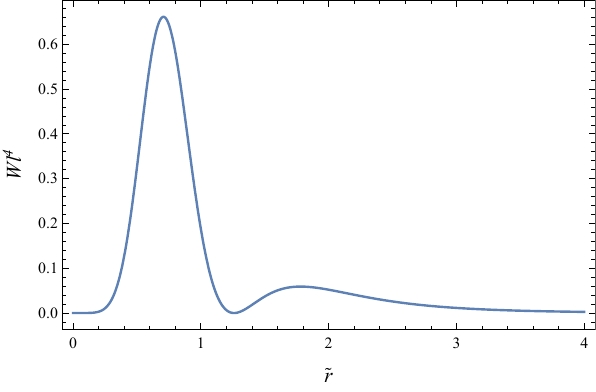}
  \end{center}
  \caption{Plot of Weyl scalar $W$ as a function of $\tilde{r}$. $W$ has two peaks and becomes zero at $\tilde{r} = 0, \sqrt[3]{2}$. The inner hgher peak is inside the outer event horizon. }
 \label{fig:Wtort}
\end{figure}

As similar as the Kretschmann scalar, the Weyl scalar can also be calculated, and the result is
\begin{equation}
 W  =  \frac{12 \tilde{r}^6 \left(\tilde{r}^3 - 2 \right)^2}{l^4 \left( \tilde{r}^3 + 1 \right)^6}.
 \label{eq:WS}
\end{equation}
Similar to the Kretschmann scalar, it can be seen that the Weyl scalar depends only on $\tilde{r}$, except for the overall factor $l^{-4}$.
The plot of the Weyl scalar is shown in Fig. \ref{fig:Wtort}.
The Weyl scalar is zero at the origin and $\tilde{r} = \sqrt[3]{2}$, and also has two maxima.
Two extreme values exist between the outer and inner horizons when the Hayward parameter $l$ is sufficiently small, whereas one exists outside the outer event horizon and the other inside the inner Cauchy horizon when $l$ is sufficiently close to the extreme value $l_0$.

\begin{figure}[ht]
 \begin{center}
   \includegraphics[width=8cm]{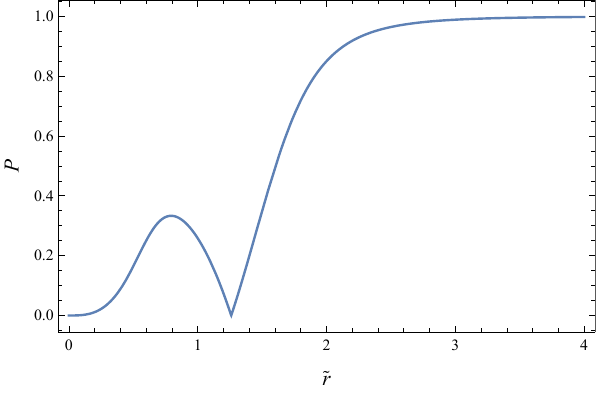}
  \end{center}
  \caption{Plot of entropic measure $P$ as a function of $\tilde{r}$. $P$ becomes zero at $\tilde{r} = \sqrt[3]{2}$ because $W=0$.}
 \label{fig:Ptort}
\end{figure}

The gravitational entropy measure can be calculated using Eqs. (\ref{eq:KS}) and (\ref{eq:WS}) to obtain
\begin{equation}
 P^2 = \frac{ \tilde{r}^6 \left( \tilde{r}^3 - 2  \right)^2 }
                  {\left( \tilde{r}^{12} - 4 \tilde{r}^9 + 18 \tilde{r}^6 - 2 \tilde{r}^3 + 2 \right) }.
 \label{eq:P_HBH}                 
\end{equation}
As can be seen from this result, the gravitational entropy measure of the Hayward black hole depends on the Hayward parameter $l$ only through $\tilde{r}$.
In Fig. \ref{fig:Ptort} we plot the entropy measure $P$ as a function of $\tilde{r}$.
Reflecting the behavior of the Weyl scalar, $P$ is zero at the origin and $\tilde{r} =  \sqrt[3]{2}$. In the region $\tilde{r} > \sqrt[3]{2}$, $P$ is a monotonically increasing function of $\tilde{r}$ and asymptotes to 1 in the limit of $\tilde{r} \rightarrow \infty$.
The limit of $\tilde{r} \rightarrow \infty$ corresponds to the limit of $l \rightarrow \infty$, i.e., the Schwarzschild black hole limit, which reproduces the fact that $P$ has a maximum value of 1 there \cite{Rudjord2008}.

\begin{figure}[ht]
 \begin{center}
   \includegraphics[width=8cm]{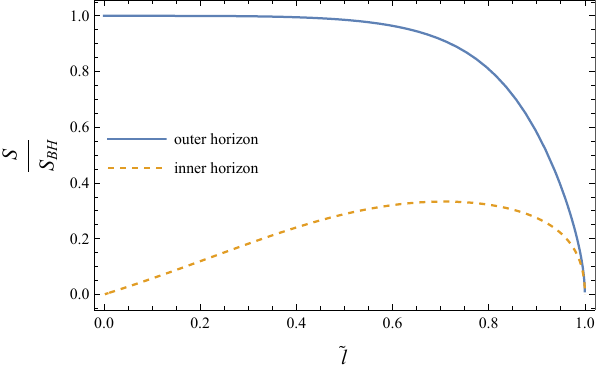}
  \end{center}
  \caption{Plot of gravitational entropy $S$ as a function of $\tilde{l}$. For extreme black hole $\tilde{l}=1$, $S=0$}
 \label{fig:SovSBHtol0}
\end{figure}

Using Eqs. (\ref{eq:def_S}), (\ref{eq:Psi}) and (\ref{eq:P_HBH}), we can calculate the gravitational entropy of the Hayward black hole.
In Fig. \ref{fig:SovSBHtol0} we show the results of our calculations.
The gravitational entropy evaluated at the outer event horizon is a monotonically decreasing function with respect to the Hayward parameter.
Since the radius of the outer event horizon $r_{+}$ is also a decreasing function of $l$, the gravitational entropy of the outer event horizon is monotonically decreasing with respect to the horizon radius.
The value of gravitational entropy evaluated at the inner Cauchy horizon has one maximum value and is zero at $l=0$ and $l_0$.
Since the Cauchy horizon radius $r_{-}$ is a monotonically increasing function with respect to $l$, the behavior of the inner horizon gravitational entropy with respect to the Cauchy horizon radius is similar to its behavior with respect to $l$ and is a decreasing function with respect to $r_{-}$ near $\tilde{r}_{-} = \sqrt[3]{2}$.
For the extreme Hayward black hole where the outer and inner horizons coincide, the gravitational entropy at the horizon is zero, the same property as for the Reisner-Nordst\"{o}m black hole~\cite{Rudjord2008}.

\begin{figure}[ht]
 \begin{center}
   \includegraphics[width=8cm]{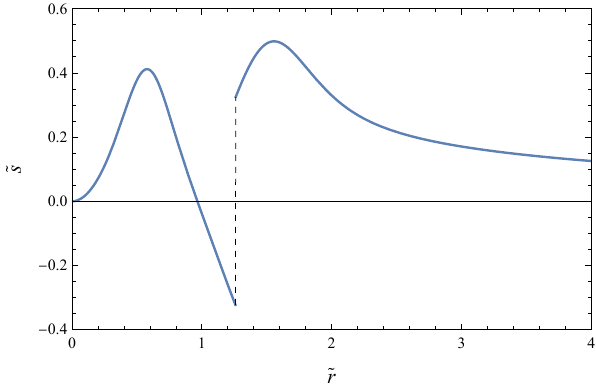}
  \end{center}
  \caption{Plot of gravitational entropy density $\tilde{s}$ as a function of $\tilde{r}$. $\tilde{s} < 0$ for $0.9 \lesssim \tilde{r} < \sqrt[3]{2}$.}
 \label{fig:s0tort}
\end{figure}

\begin{figure}[ht]
 \begin{center}
   \includegraphics[width=8cm]{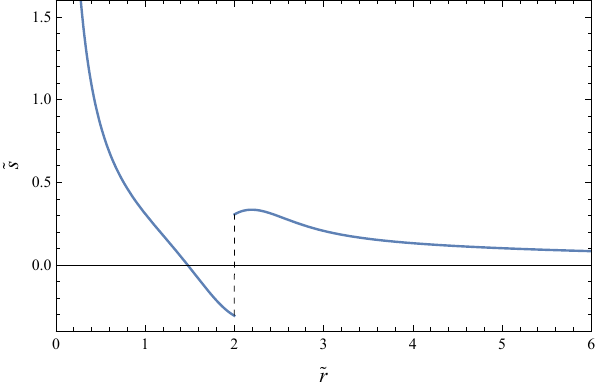}
  \end{center}
  \caption{Plot of gravitational entropy density $\tilde{s}$ of Reisner-Nordstr\"{o}m black hole as a function of $\tilde{r}$. It is similar to the Hayward black hole except that it diverges at the center.}
 \label{fig:s0tortRN}
\end{figure}

For the sake of mathematical rigor, let us perform the analysis using Eq. (\ref{eq:ent_density}) instead of Eq. (\ref{eq:abs_s}) proposed by Rudjord {\it et al.}~\cite{Rudjord2008} as the gravitational entropy density.
From the defining equation (\ref{eq:ent_density}) for the entropy density we obtain
\begin{align}
 s & =  \mathrm{sgn}(r^3 - 2 l^2 r_\mathrm{s})  k_\mathrm{S}      
       \sqrt{1 - \frac{r_\mathrm{s} r^2}{r^3 + r_\mathrm{s} l^2}}    \nonumber \\
  & \times    \left(  \frac{r^2 \left( 2 r^{15} -12 l^2 r_\mathrm{s} r^{12} + 94 l^4 r_\mathrm{s}^2 r^9 
               -85 l^6 r_\mathrm{s}^3 r^6 + 30 l^8 r_\mathrm{s}^4 r^3  -20 l^{10} r_\mathrm{s}^5 \right) } 
               {\left(r^{12} - 4 l^2 r_\mathrm{s} r^9  +18 l^4 r_\mathrm{s}^2 r^6 -2 l^6 r_\mathrm{s}^3 r^3 
                   + 2 l^8 r_\mathrm{s}^4 \right)^{\frac{3}{2}}}  \right).
\end{align}
Since $s$ is imaginary between the outer and inner horizons, we define $\tilde{s}$ using the factor $\sqrt{g_{rr}}$ in the volume element as follows
\begin{equation}
 \tilde{s} = \frac{s}{k_\mathrm{S}} \sqrt{g_{rr}} (l^2 r_\mathrm{s})^{\frac{1}{3}}.
\end{equation}
The quantity $\tilde{s}$ defined in this way does not depend explicitly on the parameters $l$ and $r_s$, and behaves as shown in Fig. \ref{fig:s0tort} as a function of $\tilde{r}$.
In the region $0.9 \lesssim \tilde{r} < \sqrt[3]{2}$, $\tilde{s}$ becomes negative.
This reflects the fact that near the extreme black hole, the gravitational entropy at the inner horizon is a decreasing function of the horizon radius $r_{-}$.
In particular, in the case of an extreme black hole with the largest inner horizon radius, the gravitational entropy of the black hole is zero as in the case of the Reisner-Nordst\"{o}m black hole, and we can conclude that the existence of a region where $s$ is negative is inevitable.
A similar analysis for the Reisner-Nordst\"{o}m black hole shows that $s$ behaves similarly for the Hayward and Reisner-Nordst\"{o}m black holes, except for the region near the center.
The Reisner-Nordstr\"{o}m black hole metric is obtained by setting $f(r)=1-\frac{rs}{r}+\frac{q^2}{r^2}$ in equation (\ref{eq:metric}), in which $q$ is the electric charge.
The gravitational entropy density behaves as shown in Fig. \ref{fig:s0tortRN}, where $\tilde{s}= \frac{s}{k_\mathrm{S}} \sqrt{g_{rr}} \frac{q^2}{r_\mathrm{s}}$ and $\tilde{r}=\frac{r_\mathrm{s}}{q^2}r$.
The elimination of the singularity at the center causes $s$ to be zero at the center for the Hayward black hole, whereas s diverges at the center for the Reisner-Nordstr\"{o}m black hole.
These results are not inconsistent with the results of Peres-Roman and Bret\'{o}n~\cite{Perez-Roman:2018hfy}, who analyzed particle motion inside the event horizon.

From a physical point of view, the entropy density should be positive, and the above results would require further investigation.
Redefining the entropy density by its absolute value, as Rudjord {\it et al.} \cite{Rudjord2008} have done, is one solution. However, it is questionable in terms of mathematical rigor.
Another possible solution is to add some correction terms to the gravitational entropy density definition equation (\ref{eq:ent_measure}).
It is shown that $0 < P^2 \leq 1$ in spherically symmetric spacetime and $P^2 = 1$ in Schwarzschild spacetime \cite{Rudjord2008}.
If $P^2=1$ can be achieved for black holes other than the Schwarzschild black hole with a correction term based on the Weyl curvature, gravitational entropy can explain the Bekenstei-Hawking entropy, and the gravitational entropy density may be positive everywhere.
If we accept that $s$ can be negative value, we cannot interpret $s$ as the gravitational entropy density.
In this case $s$ may be interpreted as the amount of change in gravitational entropy between adjacent spheres.
If so, gravitational entropy would not be defined as the integral of the vector field $\bm{\Psi}$, but as the integral of a scalar quantity $P$ on the sphere.

\section{Discussion and Summary}
\label{sec:Discussion}
In this paper, we analyze the gravitational entropy of the Hayward black hole, one of the regular black holes, based on the evaluation method of Rudjord {\it et al.}~\cite{Rudjord2008}.
In this analysis, we have written down the curvature tensor and gravitational entropy measure of the Hayward black hole, which can be in a form that brackets out the dependence on the Hayward parameter.
If we match the gravitational entropy with the Bekenstein-Hawking entropy for the Schwarschild black hole, the gravitational entropy for the Hayward black hole will be smaller than the Bekenstein-Hawking entropy and will not be proportional to the area of the event horizon.
Therefore, the gravitational entropy evaluation formula of Rudjord {\it et al.} may need some modification to make it more generally applicable to black holes.

The spacetime structure of the Hayward black hole is similar to that of the Reisner-Nordstr\"{o}m black hole, and both have an extreme black hole where the outer event horizon coincides with the inner Cauchy horizon.
In such extremal black holes, the gravitational entropy proposed by Rudjord {\it et al.} is zero.
Therefore, if we define the gravitational entropy density according to Gauss's divergence theorem, it seems inevitable that there will be regions where its value is negative.
Unlike the Reisner-Nordstr\"{o}m black hole, the gravitational entropy density of the Hayward black hole is zero at the center and does not diverge. 
We can consider this to be the result of the removal of the singularity of the Hayward black hole. Further analysis is needed to conclude whether these properties are unique to Hayward black holes or similar to other regular black holes.

It seems unavoidable that there are regions where the gravitational entropy density defined by Eq. (\ref{eq:ent_density}) is negative.
Replacing that value with an absolute value to solve this problem is one option, but not necessarily a good one.
The modification of the gravitational entropy measure (\ref{eq:ent_measure}) and the physical interpretation of the quantity defined by Eq. (\ref{eq:ent_density}) is issued to be considered in the future.

\section*{Acknowledgements}
I thank Hideki Maeda for his useful comments.


\bibliography{GE_RBH}

\begin{thebibliography}{10}
\expandafter\ifx\csname url\endcsname\relax
  \def\url#1{\texttt{#1}}\fi
\expandafter\ifx\csname urlprefix\endcsname\relax\def\urlprefix{URL }\fi
\expandafter\ifx\csname href\endcsname\relax
  \def\href#1#2{#2} \def\path#1{#1}\fi

\bibitem{LIGOScientific:2016aoc}
B.~P. Abbott, et~al., {Observation of Gravitational Waves from a Binary Black
  Hole Merger}, Phys. Rev. Lett. 116~(6) (2016) 061102.
\newblock \href {http://arxiv.org/abs/1602.03837} {\path{arXiv:1602.03837}},
  \href {https://doi.org/10.1103/PhysRevLett.116.061102}
  {\path{doi:10.1103/PhysRevLett.116.061102}}.

\bibitem{EventHorizonTelescope:2019dse}
K.~Akiyama, et~al., {First M87 Event Horizon Telescope Results. I. The Shadow
  of the Supermassive Black Hole}, Astrophys. J. Lett. 875 (2019) L1.
\newblock \href {http://arxiv.org/abs/1906.11238} {\path{arXiv:1906.11238}},
  \href {https://doi.org/10.3847/2041-8213/ab0ec7}
  {\path{doi:10.3847/2041-8213/ab0ec7}}.

\bibitem{Penrose:1964wq}
R.~Penrose, {Gravitational collapse and space-time singularities}, Phys. Rev.
  Lett. 14 (1965) 57--59.
\newblock \href {https://doi.org/10.1103/PhysRevLett.14.57}
  {\path{doi:10.1103/PhysRevLett.14.57}}.

\bibitem{Hawking:1970zqf}
S.~W. Hawking, R.~Penrose, {The Singularities of gravitational collapse and
  cosmology}, Proc. Roy. Soc. Lond. A 314 (1970) 529--548.
\newblock \href {https://doi.org/10.1098/rspa.1970.0021}
  {\path{doi:10.1098/rspa.1970.0021}}.

\bibitem{bardeen1968non}
J.~M. Bardeen, Non-singular general-relativistic gravitational collapse, in:
  Proc. Int. Conf. GR5, Tbilisi, Vol. 174, 1968, p. 174.

\bibitem{Ayon-Beato:1998hmi}
E.~Ayon-Beato, A.~Garcia, {Regular black hole in general relativity coupled to
  nonlinear electrodynamics}, Phys. Rev. Lett. 80 (1998) 5056--5059.
\newblock \href {http://arxiv.org/abs/gr-qc/9911046}
  {\path{arXiv:gr-qc/9911046}}, \href
  {https://doi.org/10.1103/PhysRevLett.80.5056}
  {\path{doi:10.1103/PhysRevLett.80.5056}}.

\bibitem{Ayon-Beato:1999qin}
E.~Ayon-Beato, A.~Garcia, {Nonsingular charged black hole solution for
  nonlinear source}, Gen. Rel. Grav. 31 (1999) 629--633.
\newblock \href {http://arxiv.org/abs/gr-qc/9911084}
  {\path{arXiv:gr-qc/9911084}}, \href {https://doi.org/10.1023/A:1026640911319}
  {\path{doi:10.1023/A:1026640911319}}.

\bibitem{Ayon-Beato:1999kuh}
E.~Ayon-Beato, A.~Garcia, {New regular black hole solution from nonlinear
  electrodynamics}, Phys. Lett. B 464 (1999) 25.
\newblock \href {http://arxiv.org/abs/hep-th/9911174}
  {\path{arXiv:hep-th/9911174}}, \href
  {https://doi.org/10.1016/S0370-2693(99)01038-2}
  {\path{doi:10.1016/S0370-2693(99)01038-2}}.

\bibitem{Ayon-Beato:2000mjt}
E.~Ayon-Beato, A.~Garcia, {The Bardeen model as a nonlinear magnetic monopole},
  Phys. Lett. B 493 (2000) 149--152.
\newblock \href {http://arxiv.org/abs/gr-qc/0009077}
  {\path{arXiv:gr-qc/0009077}}, \href
  {https://doi.org/10.1016/S0370-2693(00)01125-4}
  {\path{doi:10.1016/S0370-2693(00)01125-4}}.

\bibitem{Ayon-Beato:2004ywd}
E.~Ayon-Beato, A.~Garcia, {Four parametric regular black hole solution}, Gen.
  Rel. Grav. 37 (2005) 635.
\newblock \href {http://arxiv.org/abs/hep-th/0403229}
  {\path{arXiv:hep-th/0403229}}, \href
  {https://doi.org/10.1007/s10714-005-0050-y}
  {\path{doi:10.1007/s10714-005-0050-y}}.

\bibitem{Dymnikova:1992ux}
I.~Dymnikova, {Vacuum nonsingular black hole}, Gen. Rel. Grav. 24 (1992)
  235--242.
\newblock \href {https://doi.org/10.1007/BF00760226}
  {\path{doi:10.1007/BF00760226}}.

\bibitem{Bronnikov:1979ex}
K.~A. Bronnikov, V.~N. Melnikov, G.~N. Shikin, K.~P. Staniukowicz, {SCALAR,
  ELECTROMAGNETIC, AND GRAVITATIONAL FIELDS INTERACTION: PARTICLE - LIKE
  SOLUTIONS}, Annals Phys. 118 (1979) 84--107.
\newblock \href {https://doi.org/10.1016/0003-4916(79)90235-5}
  {\path{doi:10.1016/0003-4916(79)90235-5}}.

\bibitem{Bronnikov:2000vy}
K.~A. Bronnikov, {Regular magnetic black holes and monopoles from nonlinear
  electrodynamics}, Phys. Rev. D 63 (2001) 044005.
\newblock \href {http://arxiv.org/abs/gr-qc/0006014}
  {\path{arXiv:gr-qc/0006014}}, \href
  {https://doi.org/10.1103/PhysRevD.63.044005}
  {\path{doi:10.1103/PhysRevD.63.044005}}.

\bibitem{Bronnikov:2022ofk}
K.~A. Bronnikov, {Regular black holes sourced by nonlinear electrodynamics} (11
  2022).
\newblock \href {http://arxiv.org/abs/2211.00743} {\path{arXiv:2211.00743}}.

\bibitem{Fan2016}
Z.-Y. Fan, X.~Wang, \href{http://arxiv.org/abs/1610.02636
  http://dx.doi.org/10.1103/PhysRevD.94.124027}{Construction of regular black
  holes in general relativity}, Physical Review D 94 (2016) 124027.
\newblock \href {https://doi.org/10.1103/PhysRevD.94.124027}
  {\path{doi:10.1103/PhysRevD.94.124027}}.
\newline\urlprefix\url{http://arxiv.org/abs/1610.02636
  http://dx.doi.org/10.1103/PhysRevD.94.124027}

\bibitem{Hayward2006}
S.~A. Hayward, Formation and evaporation of nonsingular black holes, Physical
  Review Letters 96 (1 2006).
\newblock \href {https://doi.org/10.1103/PhysRevLett.96.031103}
  {\path{doi:10.1103/PhysRevLett.96.031103}}.

\bibitem{Ansoldi:2008jw}
S.~Ansoldi, {Spherical black holes with regular center: A Review of existing
  models including a recent realization with Gaussian sources}, in: {Conference
  on Black Holes and Naked Singularities}, 2008.
\newblock \href {http://arxiv.org/abs/0802.0330} {\path{arXiv:0802.0330}}.

\bibitem{Lemos:2011dq}
J.~P.~S. Lemos, V.~T. Zanchin, {Regular black holes: Electrically charged
  solutions, Reissner-Nordstr\"om outside a de Sitter core}, Phys. Rev. D 83
  (2011) 124005.
\newblock \href {http://arxiv.org/abs/1104.4790} {\path{arXiv:1104.4790}},
  \href {https://doi.org/10.1103/PhysRevD.83.124005}
  {\path{doi:10.1103/PhysRevD.83.124005}}.

\bibitem{Maeda2022}
H.~Maeda, \href{https://link.springer.com/10.1007/JHEP11(2022)108}{Quest for
  realistic non-singular black-hole geometries: regular-center type}, Journal
  of High Energy Physics 2022 (2022) 108.
\newblock \href {https://doi.org/10.1007/JHEP11(2022)108}
  {\path{doi:10.1007/JHEP11(2022)108}}.
\newline\urlprefix\url{https://link.springer.com/10.1007/JHEP11(2022)108}

\bibitem{Isomura:2023oqf}
K.~Isomura, R.~Suzuki, S.~Tomizawa, {Particle motions around regular black
  holes} (1 2023).
\newblock \href {http://arxiv.org/abs/2301.10465} {\path{arXiv:2301.10465}}.

\bibitem{penrose1979singularities}
R.~{Penrose}, {Singularities and time-asymmetry.}, in: S.~W. {Hawking},
  W.~{Israel} (Eds.), General Relativity: An Einstein centenary survey, 1979,
  pp. 581--638.

\bibitem{Rudjord2008}
{\O}.~Rudjord, {\O}.~Gr{\o}n, S.~Hervik, The weyl curvature conjecture and
  black hole entropy, Physica Scripta 77 (5 2008).
\newblock \href {https://doi.org/10.1088/0031-8949/77/05/055901}
  {\path{doi:10.1088/0031-8949/77/05/055901}}.

\bibitem{Gron:2012}
{\O}.~Gr{\o}n, \href{http://www.mdpi.com/1099-4300/14/12/2456}{Entropy and
  gravity}, Entropy 14 (2012) 2456--2477.
\newblock \href {https://doi.org/10.3390/e14122456}
  {\path{doi:10.3390/e14122456}}.
\newline\urlprefix\url{http://www.mdpi.com/1099-4300/14/12/2456}

\bibitem{Romero2012}
G.~E. Romero, R.~Thomas, D.~P{\'e}rez,
  \href{http://link.springer.com/10.1007/s10773-011-0967-8}{Gravitational
  entropy of black holes and wormholes}, International Journal of Theoretical
  Physics 51 (2012) 925--942.
\newblock \href {https://doi.org/10.1007/s10773-011-0967-8}
  {\path{doi:10.1007/s10773-011-0967-8}}.
\newline\urlprefix\url{http://link.springer.com/10.1007/s10773-011-0967-8}

\bibitem{Perez:2014oea}
D.~P\'erez, G.~E. Romero, S.~E. Perez-Bergliaffa, {An Analysis of a Regular
  Black Hole Interior Model}, Int. J. Theor. Phys. 53 (2014) 734--753.
\newblock \href {https://doi.org/10.1007/s10773-013-1861-3}
  {\path{doi:10.1007/s10773-013-1861-3}}.

\bibitem{Guha:2019fun}
S.~Guha, S.~Chakraborty, {On the gravitational entropy of accelerating black
  holes}, Int. J. Mod. Phys. D 29~(05) (2020) 2050034.
\newblock \href {http://arxiv.org/abs/1908.06763} {\path{arXiv:1908.06763}},
  \href {https://doi.org/10.1142/S0218271820500340}
  {\path{doi:10.1142/S0218271820500340}}.

\bibitem{deCLima:2020rvr}
R.~de~C.~Lima, J.~A.~C. Nogales, S.~H. Pereira, {Gravitational entropy of
  wormholes with exotic matter and in galactic halos}, Int. J. Mod. Phys. D
  29~(02) (2020) 2050015.
\newblock \href {http://arxiv.org/abs/2003.08344} {\path{arXiv:2003.08344}},
  \href {https://doi.org/10.1142/S0218271820500157}
  {\path{doi:10.1142/S0218271820500157}}.

\bibitem{Chakraborty2022}
S.~Chakraborty, S.~Guha, R.~Goswami,
  \href{https://link.springer.com/10.1007/s10714-022-02934-3}{How appropriate
  are the gravitational entropy proposals for traversable wormholes?}, General
  Relativity and Gravitation 54 (2022) 47.
\newblock \href {https://doi.org/10.1007/s10714-022-02934-3}
  {\path{doi:10.1007/s10714-022-02934-3}}.
\newline\urlprefix\url{https://link.springer.com/10.1007/s10714-022-02934-3}

\bibitem{Clifton2013}
T.~Clifton, G.~F.~R. Ellis, R.~Tavakol,
  \href{https://iopscience.iop.org/article/10.1088/0264-9381/30/12/125009}{A
  gravitational entropy proposal}, Classical and Quantum Gravity 30 (2013)
  125009.
\newblock \href {https://doi.org/10.1088/0264-9381/30/12/125009}
  {\path{doi:10.1088/0264-9381/30/12/125009}}.
\newline\urlprefix\url{https://iopscience.iop.org/article/10.1088/0264-9381/30/12/125009}

\bibitem{Gregoris2022}
D.~Gregoris, Y.~C. Ong, Understanding gravitational entropy of black holes: A
  new proposal via curvature invariants, Physical Review D 105 (5 2022).
\newblock \href {https://doi.org/10.1103/PhysRevD.105.104017}
  {\path{doi:10.1103/PhysRevD.105.104017}}.

\bibitem{Mbonye:2005im}
M.~R. Mbonye, D.~Kazanas, {A Non-singular black hole model as a possible
  end-product of gravitational collapse}, Phys. Rev. D 72 (2005) 024016.
\newblock \href {http://arxiv.org/abs/gr-qc/0506111}
  {\path{arXiv:gr-qc/0506111}}, \href
  {https://doi.org/10.1103/PhysRevD.72.024016}
  {\path{doi:10.1103/PhysRevD.72.024016}}.

\bibitem{Bekenstein:1973ur}
J.~D. Bekenstein, {Black holes and entropy}, Phys. Rev. D 7 (1973) 2333--2346.
\newblock \href {https://doi.org/10.1103/PhysRevD.7.2333}
  {\path{doi:10.1103/PhysRevD.7.2333}}.

\bibitem{Bardeen:1973gs}
J.~M. Bardeen, B.~Carter, S.~W. Hawking, {The Four laws of black hole
  mechanics}, Commun. Math. Phys. 31 (1973) 161--170.
\newblock \href {https://doi.org/10.1007/BF01645742}
  {\path{doi:10.1007/BF01645742}}.

\bibitem{Hawking:1975vcx}
S.~W. Hawking, {Particle Creation by Black Holes}, Commun. Math. Phys. 43
  (1975) 199--220, [Erratum: Commun.Math.Phys. 46, 206 (1976)].
\newblock \href {https://doi.org/10.1007/BF02345020}
  {\path{doi:10.1007/BF02345020}}.

\bibitem{Hawking:1976de}
S.~W. Hawking, {Black Holes and Thermodynamics}, Phys. Rev. D 13 (1976)
  191--197.
\newblock \href {https://doi.org/10.1103/PhysRevD.13.191}
  {\path{doi:10.1103/PhysRevD.13.191}}.

\bibitem{Chinaglia:2017uqd}
S.~Chinaglia, S.~Zerbini, {A note on singular and non-singular black holes},
  Gen. Rel. Grav. 49~(6) (2017) 75.
\newblock \href {http://arxiv.org/abs/1704.08516} {\path{arXiv:1704.08516}},
  \href {https://doi.org/10.1007/s10714-017-2235-6}
  {\path{doi:10.1007/s10714-017-2235-6}}.

\bibitem{Chinaglia:2018gvf}
S.~Chinaglia, {A no-go theorem for regular black holes} (5 2018).
\newblock \href {http://arxiv.org/abs/1805.03899} {\path{arXiv:1805.03899}}.

\bibitem{Toshmatov:2018cks}
B.~Toshmatov, Z.~Stuchl\'\i{}k, B.~Ahmedov, {Comment on
  \textquotedblleft{}Construction of regular black holes in general
  relativity\textquotedblright{}}, Phys. Rev. D 98~(2) (2018) 028501.
\newblock \href {http://arxiv.org/abs/1807.09502} {\path{arXiv:1807.09502}},
  \href {https://doi.org/10.1103/PhysRevD.98.028501}
  {\path{doi:10.1103/PhysRevD.98.028501}}.

\bibitem{Tharanath:2014naa}
R.~Tharanath, J.~Suresh, V.~C. Kuriakose, {Phase transitions and
  Geometrothermodynamics of Regular black holes}, Gen. Rel. Grav. 47~(4) (2015)
  46.
\newblock \href {http://arxiv.org/abs/1406.3916} {\path{arXiv:1406.3916}},
  \href {https://doi.org/10.1007/s10714-015-1884-6}
  {\path{doi:10.1007/s10714-015-1884-6}}.

\bibitem{Maluf2018}
R.~Maluf, J.~C. Neves,
  \href{https://link.aps.org/doi/10.1103/PhysRevD.97.104015}{Thermodynamics of
  a class of regular black holes with a generalized uncertainty principle},
  Physical Review D 97 (2018) 104015.
\newblock \href {https://doi.org/10.1103/PhysRevD.97.104015}
  {\path{doi:10.1103/PhysRevD.97.104015}}.
\newline\urlprefix\url{https://link.aps.org/doi/10.1103/PhysRevD.97.104015}

\bibitem{Molina:2021hgx}
M.~Molina, J.~R. Villanueva, {On the thermodynamics of the Hayward black hole},
  Class. Quant. Grav. 38~(10) (2021) 105002.
\newblock \href {http://arxiv.org/abs/2101.07917} {\path{arXiv:2101.07917}},
  \href {https://doi.org/10.1088/1361-6382/abdd47}
  {\path{doi:10.1088/1361-6382/abdd47}}.

\bibitem{Fathi2021}
M.~Fathi, M.~Molina, J.~R. Villanueva, Adiabatic evolution of hayward black
  hole, Physics Letters, Section B: Nuclear, Elementary Particle and
  High-Energy Physics 820 (9 2021).
\newblock \href {https://doi.org/10.1016/j.physletb.2021.136548}
  {\path{doi:10.1016/j.physletb.2021.136548}}.

\bibitem{IlichKruglov:2021pdw}
S.~Il'ich~Kruglov, {Remarks on Nonsingular Models of Hayward and Magnetized
  Black Hole with Rational Nonlinear Electrodynamics}, Grav. Cosmol. 27~(1)
  (2021) 78--84.
\newblock \href {http://arxiv.org/abs/2103.14087} {\path{arXiv:2103.14087}},
  \href {https://doi.org/10.1134/S0202289321010126}
  {\path{doi:10.1134/S0202289321010126}}.

\bibitem{Ma2014}
M.-S. Ma, R.~Zhao, \href{http://arxiv.org/abs/1411.0833
  http://dx.doi.org/10.1088/0264-9381/31/24/245014}{Corrected form of the first
  law of thermodynamics for regular black holes}, Classical and Quantum Gravity
  31 (11 2014).
\newblock \href {https://doi.org/10.1088/0264-9381/31/24/245014}
  {\path{doi:10.1088/0264-9381/31/24/245014}}.
\newline\urlprefix\url{http://arxiv.org/abs/1411.0833
  http://dx.doi.org/10.1088/0264-9381/31/24/245014}

\bibitem{Ma:2015gpa}
M.-S. Ma, {Magnetically charged regular black hole in a model of nonlinear
  electrodynamics}, Annals Phys. 362 (2015) 529--537.
\newblock \href {http://arxiv.org/abs/1509.05580} {\path{arXiv:1509.05580}},
  \href {https://doi.org/10.1016/j.aop.2015.08.028}
  {\path{doi:10.1016/j.aop.2015.08.028}}.

\bibitem{Zhang:2016ilt}
Y.~Zhang, S.~Gao, {First law and Smarr formula of black hole mechanics in
  nonlinear gauge theories}, Class. Quant. Grav. 35~(14) (2018) 145007.
\newblock \href {http://arxiv.org/abs/1610.01237} {\path{arXiv:1610.01237}},
  \href {https://doi.org/10.1088/1361-6382/aac9d4}
  {\path{doi:10.1088/1361-6382/aac9d4}}.

\bibitem{Rasheed:1997ns}
D.~A. Rasheed, {Nonlinear electrodynamics: Zeroth and first laws of black hole
  mechanics} (2 1997).
\newblock \href {http://arxiv.org/abs/hep-th/9702087}
  {\path{arXiv:hep-th/9702087}}.

\bibitem{Fan:2016rih}
Z.-Y. Fan, {Critical phenomena of regular black holes in anti-de Sitter
  space-time}, Eur. Phys. J. C 77~(4) (2017) 266.
\newblock \href {http://arxiv.org/abs/1609.04489} {\path{arXiv:1609.04489}},
  \href {https://doi.org/10.1140/epjc/s10052-017-4830-9}
  {\path{doi:10.1140/epjc/s10052-017-4830-9}}.

\bibitem{Bronnikov:2017tnz}
K.~A. Bronnikov, {Comment on \textquotedblleft{}Construction of regular black
  holes in general relativity\textquotedblright{}}, Phys. Rev. D 96~(12) (2017)
  128501.
\newblock \href {http://arxiv.org/abs/1712.04342} {\path{arXiv:1712.04342}},
  \href {https://doi.org/10.1103/PhysRevD.96.128501}
  {\path{doi:10.1103/PhysRevD.96.128501}}.

\bibitem{Chiba:2017nml}
T.~Chiba, M.~Kimura, \href{https://doi.org/10.1093/ptep/ptx037}{{A note on
  geodesics in the Hayward metric}}, Progress of Theoretical and Experimental
  Physics 2017~(4), 043E01 (04 2017).
\newblock \href {https://doi.org/10.1093/ptep/ptx037}
  {\path{doi:10.1093/ptep/ptx037}}.
\newline\urlprefix\url{https://doi.org/10.1093/ptep/ptx037}

\bibitem{Perez-Roman:2018hfy}
I.~Perez-Roman, N.~Bret\'on, {The region interior to the event horizon of the
  Regular Hayward Black Hole}, Gen. Rel. Grav. 50~(6) (2018) 64.
\newblock \href {http://arxiv.org/abs/1805.00906} {\path{arXiv:1805.00906}},
  \href {https://doi.org/10.1007/s10714-018-2385-1}
  {\path{doi:10.1007/s10714-018-2385-1}}.

\end{thebibliography}
\bibliographystyle{elsarticle-num}

\end{document}